\documentclass[aps,prd,notitlepage,twocolumn, superscriptaddress, nolongbibliography, 10pt]{revtex4-2}

\usepackage{graphicx} 
\usepackage[utf8]{inputenc}
\usepackage{amssymb,amsmath,amsfonts}
\usepackage{multirow}
\usepackage{slashed}
\usepackage{graphicx}
\usepackage{subfigure}
\usepackage{hyperref}
\hypersetup{colorlinks=true,anchorcolor=pink,citecolor=cyan, filecolor=blue,urlcolor=red,bookmarksnumbered=true,
	pdfview=FitB
}
\usepackage{cleveref}

\usepackage{xcolor}
\definecolor{XHCmathematicaTitleRed}{RGB}{204,11,2}

\begin{document}

\title{Nuclear structure functions and EMC effect in a relativistic deuteron}
\author{Xianghui Cao}
\affiliation{Department of Modern Physics, University of Science and Technology of China, Hefei 230026, China}

\author{Bheemsehan Gurjar}
\affiliation{Department of Modern Physics, University of Science and Technology of China, Hefei 230026, China}

\author{Vladimir~A.~Karmanov}
\affiliation{Lebedev Physical Institute, Leninsky Prospekt 53, 119991 Moscow, Russia}

\author{Yang Li}
\affiliation{Department of Modern Physics, University of Science and Technology of China, Hefei 230026, China}
\affiliation{Anhui Center for Fundamental Sciences in Theoretical Physics, Hefei, 230026, China}

\date{\today}

\begin{abstract}
    We investigate the longitudinal momentum distributions (LMDs) of the deuteron using a relativistic deuteron wave function derived from a realistic nucleon-nucleon (NN) interaction within the light-front framework. By convoluting the deuteron LMDs with the nucleon parton distribution functions (PDFs), we analyze the unpolarized, longitudinally polarized, and tensor-polarized structure functions of the deuteron. In addition, we also study the EMC effect in the weakly bound deuteron through the structure function ratio $F_{2}^{D}/(F_{2}^{p}+F_{2}^{n})$, which deviates from unity due to nuclear binding and Fermi motion effects.
\end{abstract}

\maketitle 
\section{Introduction}\label{sec:Introduction}
The spin structure of spin-$1/2$ systems has been investigated extensively through measurements of polarized structure functions in Deep Inelastic Scattering (DIS) experiments~\cite{EuropeanMuon:1987isl}. However, a complete understanding of the nucleon spin structure in terms of its constituent quarks and gluons remains elusive~\cite{Ji:2020ena}. In particular, the gluon spin contribution and the orbital angular momentum (OAM) of quarks and gluons are still not well constrained experimentally. These fundamental aspects of nucleon structure are among the primary objectives of future studies at upcoming Electron-Ion Collider (EIC) facilities~\cite{AbdulKhalek:2021gbh,Anderle:2021wcy}. On the other hand, spin-$1$ hadrons possess a richer spin structure than spin-$1/2$ nucleons. However, our understanding of their spin structure is still in early stages due to the lack of experimental measurements of polarized deuteron structure functions since the HERMES $b_{1}$-experiment~\cite{HERMES:2005pon}. Although several experimental programs are planned to shed light on the spin structure of spin-$1$ hadrons, particularly the deuteron, at Jefferson Lab~\cite{Maxwell:2018gci,ProposalJLab2023}, Fermilab~\cite{Keller:2022abm,AbilityFermiLabE1039}, Nuclotron-based Ion Collider fAcility (NICA)~\cite{Arbuzov:2020cqg}, and future Electron-Ion Collider facilities~\cite{AbdulKhalek:2021gbh,Anderle:2021wcy}.

If the deuteron were simply an S-wave bound state of a proton and a neutron, the tensor polarized structure function $b_{1}$ would vanish. Even after including the conventional D-state admixture in the standard convolution model, theoretical calculations predict $b_{1}$ which differs significantly from the measurements reported by the HERMES collaboration~\cite{Cosyn:2017fbo,Kumano:2024fpr,Hoodbhoy:1988am,Khan:1991qk}. The observed nonzero $b_{1}$ therefore suggests that the deuteron possesses a more complex internal structure beyond the conventional proton-neutron picture~\cite{Machleidt:1987hj}. One possible explanation is the presence of hidden-color configurations arising from six-quark components in the deuteron wave function, which could provide a significant contribution to the measured tensor structure function $b_{1}$ and indicate the need for new hadronic mechanisms~\cite{Ji:2014cda,Miller:2013hla,Kaur:2025css}. The proposals at JLab, Fermilab, NICA and EICs also aim to clarify the discrepancy between the existing theoretical predictions through the standard convolution description and the available HERMES data for the tensor-polarized structure function $b_{1}$ (see the recent review~\cite{Poudel:2025nof} for updated information). On the other hand, the unpolarized deuteron structure function $F_{2}$ and the longitudinally polarized structure function $g_{1}$ have been extensively measured in experiments by the New Muon Collaboration (NMC)~\cite{NewMuon:1995zeb,NewMuon:1996fwh}, the Spin Muon Collaboration (SMC)~\cite{SpinMuon:1993gcv,SpinMuon:1995svc,SpinMuon:1998eqa,SpinMuon:1999udj}, HERMES~\cite{HERMES:2006jyl}, and COMPASS~\cite{COMPASS:2005xxc} collaborations, respectively.

Furthermore, nuclear structure functions provide an essential baseline for investigating nuclear modifications of the partonic structure of bound protons and neutrons, commonly referred to as the ``nuclear EMC effect"~\cite{EuropeanMuon:1983wih}. This phenomenon is characterized by significant deviations of the ratio of nuclear to isoscalar nucleon $(N=p+n)$ structure functions from unity, reflecting modifications of nucleon parton distributions in the nuclear medium~\cite{Bodek:1983qn}. Owing to its weak binding energy and simple two-nucleon structure, the deuteron exhibits only minimal nuclear modifications compared to heavier nuclei, making it an ideal reference for isolating genuine medium effects~\cite{Seely:2009gt,Cocuzza:2026vey}.

While the bulk properties of the deuteron at low energies are successfully described in terms of the conventional nonrelativistic $S$- and $D$-wave components~\cite{Garcon:2001sz,Machleidt:2000ge,Wiringa:1994wb}, a reliable description of its high-energy structure requires a relativistic treatment of its internal dynamics. High-energy scattering processes can become sensitive to short-distance nuclear configurations and to the high-momentum components of the deuteron wave function, where relativistic effects are expected to be significant~\cite{Frankfurt:1981mk, Frankfurt:1988nt, Arrington:2022sov}. A consistent treatment beyond the nonrelativistic regime is therefore essential for connecting conventional nuclear dynamics with observables measured in high-energy reactions.

Light-front dynamics (LFD) provides a natural relativistic framework for studying nuclear structure in high-energy scattering processes~\cite{Brodsky:1997de, Miller:2000kv, Vary:2025yqo}. Unlike conventional instant-form approaches, LFD allows a clean separation between internal and center-of-mass motion, while the longitudinal boosts remain kinematical. Consequently, the internal light-front wave functions (LFWFs) are directly connected to the momentum variables entering high-energy factorization and convolution approaches~\cite{Lepage:1980fj, Cosyn:2017fbo}. This makes LFD particularly suitable for describing the partonic structure of the deuteron and studying observables such as nuclear structure functions and the EMC effect~\cite{Carbonell:1995yi, Cooke:2001kz, Karmanov:2025lod, Zhu:2026iyk}.

A widely used strategy in light-front studies of nuclear structure is to construct the nuclear LFWFs from conventional nonrelativistic wave functions by applying the Melosh rotation or a variable substitution~\cite{Frankfurt:1993ut,Frederico:1991vb,Kondratyuk:1983kq,Chung:1988my,Kamano:2012yx,Ehlers:2014jpa,Pace:2016eiq,Pace:2022qoj,Fornetti:2023gvf,Proietti:2024bhb,Cosyn:2017fbo,Cosyn:2019hem,Cosyn:2020kwu,Segarra:2020plg}. This approach maintains both the successful low-energy description supplied by high-precision nuclear potentials and the relativistic nature of LFD. It has been extensively employed in calculations of deuteron form factors~\cite{Frankfurt:1993ut,Frederico:1991vb,Kondratyuk:1983kq,Chung:1988my}, the EMC effect~\cite{Pace:2016eiq,Pace:2022qoj,Fornetti:2023gvf,Segarra:2020plg}, the tagged deep-inelastic scattering~\cite{Cosyn:2019hem,Cosyn:2020kwu}, structure functions~\cite{Proietti:2024bhb,Cosyn:2017fbo}, and the sea-quark asymmetry~\cite{Kamano:2012yx,Ehlers:2014jpa}. Although such constructions provide a practical connection between conventional nuclear wave functions and LFD, the underlying interactions are generally imported from an instant-form or nonrelativistic calculation. A more straightforward methodology is to construct the NN interaction directly within LFD and solve the corresponding relativistic bound-state equation. Earlier studies following this direction have investigated only conventional observables such as the binding energy and electromagnetic form factors~\cite{Karmanov:1991fv,Karmanov:1994ck,Carbonell:1999pt,Cooke:2001kz}.

In this work, we employ relativistic deuteron wave functions based on a realistic NN interaction within the light-front framework~\cite{Karmanov:1994ck,Machleidt:1987hj}.
The underlying NN potential provides a quantitative description of low-energy NN scattering data and has been widely applied to a variety of nuclear observables, including electron-deuteron scattering~\cite{Pauschenwein:1990ljs,Witala:1989pxg,Arenhovel:1999nq}, deuteron photo- and electrodisintegration~\cite{Schmitt:1990yb,Ritz:1996za}, the ${}^3 \text{H}$ binding energy~\cite{Stadler:1995wu}, and the electromagnetic form factors of ${}^3\text{He} / {}^3 \text{H}$~\cite{Pena:1990ea,Henning:1992tq}.
Within the light-front formulation, the deuteron LFWFs are represented in terms of six invariant scalar functions, rather than the non-relativistic S- and D-wave components~\cite{Karmanov:1980mc,Carbonell:1995yi}. The same LFWFs have been utilized to predict the deuteron electromagnetic form factors~\cite{Karmanov:1991fv,Karmanov:1994ck,Carbonell:1999pt}, yielding a good description of the experimental data over a broad range of momentum transfers , up to $Q^2 \lesssim 3 ~\text{GeV}^2$ (a comprehensive discussion of this framework is provided in the review~\cite{Carbonell:1998rj}).
We then use the overlap representation of deuteron LFWFs to compute the deuteron 
LMDs for unpolarized, longitudinally polarized, and tensor-polarized deuteron configurations, along with the corresponding structure functions~\cite{Bacchetta:2000jk,Frankfurt:1981mk,Hoodbhoy:1988am}. These structure functions are evaluated within the standard convolution formalism, in which the deuteron LMDs are convoluted with the nucleon parton distribution functions adopted from the neural-network-based parton distribution function (NNPDF) collaboration~\cite{NNPDF:2017mvq}.
Furthermore, we investigate the deuteron EMC effect by calculating the ratio of the deuteron unpolarized structure function to that of the free proton and neutron~\cite{NuSea:1998kqi,NuSea:2001idv,SeaQuest:2021zxb}.
We find that the overall predictions of our analysis are consistent with existing phenomenological and theoretical studies, as well as with experimental measurements reported in the literature.

\begin{figure*}
    \centering
    \includegraphics[scale=0.5]{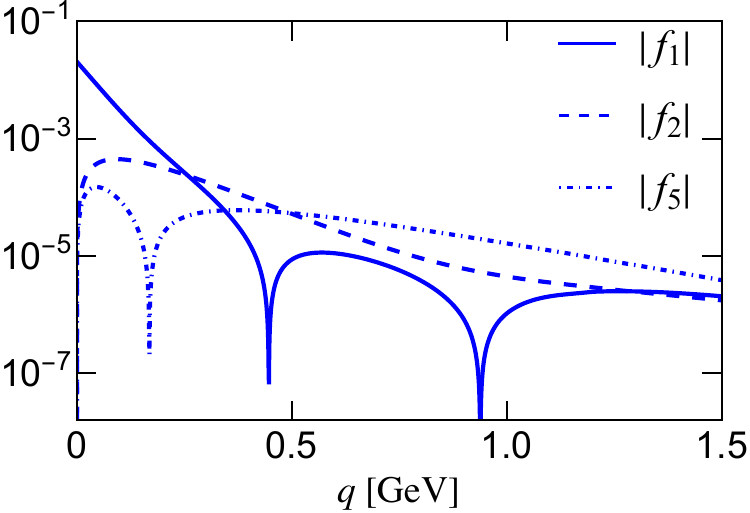}~~~\hspace{1cm}
    \includegraphics[scale=0.5]{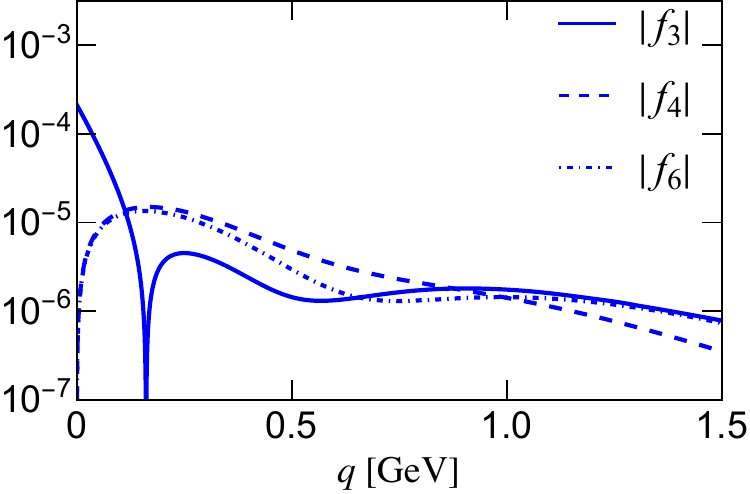}
    \caption{The scalar components $f_{1-6}$ of Eq.~\eqref{eq:f1tof6} in the decomposition of the deuteron LFWFs versus $q$ at fixed $z = \vec{\omega} \cdot \vec{q} / |\vec{q}| =0.6$.}
    \label{fig:Deuteron-LFWF}
\end{figure*}
%
\section{Deuteron light-front wave function}\label{sec:model}
The relativistic LFWFs of a spin-1 deuteron with four-momentum $p$ and polarization  $\lambda$, composed of two spin-1/2 on mass shell nucleons with spin $\sigma_{1}$ and $\sigma_{2}$, are expressed as~\cite{Carbonell:1995yi},
\begin{gather}
    \Psi_{\sigma_2\sigma_1}^\lambda = e_\mu^{(\lambda)}(p) \bar{u}^{\sigma_2}(k_2) \phi^\mu U_c \left(\bar{u}^{\sigma_1}\right)^{T}(k_1)\,.
    \label{eq:deuteron-LFWF-spinor-structure}
\end{gather}
where $k_{1}$ and $k_{2}$ are the mass shell four-momentum of the constituent nucleons, $U_c = \gamma^2\gamma^0$ is the charge conjugation operator, $\bar{u}$ and $u$ denotes the nucleon four-component Dirac spinors, and $e_\mu^{(\lambda)}(p)$ is the deuteron polarization vector.
The tensor $\phi^\mu$ admits a decomposition into six scalar functions on the light-front as,
\begin{multline}
    \phi^\mu = \varphi_1 \frac{(k_1 - k_2)^\mu}{2m^2} + \varphi_2 \frac{1}{m}\gamma^\mu + \varphi_3 \frac{\omega^\mu}{\omega \cdot p} + \varphi_4 \frac{(k_1-k_2)^\mu \slashed{\omega}}{2m(\omega \cdot p)}\\
    + \varphi_5 \frac{i \gamma_5 \epsilon^{\mu\nu\rho\gamma}(k_1 + k_2)_\nu (k_1 - k_2)_\rho \omega_\gamma}{2m^2 (\omega \cdot p)} + \varphi_6 \frac{m \omega^\mu \slashed{\omega}}{(\omega \cdot p)^2}\,.
    \label{eq:deuteron-LFWF-tensor-structure}
\end{multline}
where $\epsilon^{\mu\nu\rho\gamma}$ denotes the Levi-Civita tensor, $\gamma^{\mu}$ and $\gamma_5$ are the Dirac matrices, and $m$ is the nucleon mass. The null vector $\omega^\mu$ (with $\omega^2 = 0$) specifies the orientation of the light front.
\Cref{eq:deuteron-LFWF-spinor-structure,eq:deuteron-LFWF-tensor-structure} are fully covariant. It is convenient to construct the LFWFs in terms of the particle momentum variables $\vec{q}$ and $\vec{n}$, which coincide with $\vec{k}_1$ and $\vec{\omega} / |\vec{\omega}|$ in the center-of-mass frame $\vec{k}_1 + \vec{k}_2 = 0$ (see Eqs.~(3.11), (3.12) of~\cite{Carbonell:1998rj}).
Then the deuteron LFWF reduces to,
\begin{align}
    \Psi_{\sigma_2\sigma_1}^\lambda &= w^*_{\sigma_2} \psi^\lambda(\vec{q}, \vec{\omega}) \sigma_y \left(w_{\sigma_1}^{*}\right)^{T}\,,
\end{align}
with,
\begin{align}\label{eq:f1tof6}
    \vec{\psi}(\vec{q}, \vec{n}) &= f_1 \frac{\vec{\sigma}}{\sqrt{2}} + f_2 \frac{1}{2} \left( \frac{3\vec{q}(\vec{q} \cdot \vec{\sigma})}{\vec{q}^2} - \vec{\sigma} \right) \notag\\
    &+ f_3 \frac{1}{2} (3 \vec{n} (\vec{n} \cdot \vec{\sigma}) - \vec{\sigma} )
    \notag\\
    &+ f_4 \frac{1}{2|\vec{q}|} (3\vec{q} (\vec{n} \cdot \vec{\sigma}) + 3 \vec{n} (\vec{q} \cdot \vec{\sigma}) - 2(\vec{q} \cdot \vec{n}) \vec{\sigma} )
 \notag\\
    &+ f_5 \sqrt{\frac{3}{2}} \frac{i}{|\vec{q}|} [\vec{q} \times \vec{n}] + f_6 \frac{\sqrt{3}}{2|\vec{q}|}[[\vec{q} \times \vec{n}] \times \vec{\sigma}]\,.
\end{align}
Here $\vec{\sigma}$ denotes the Pauli matrices.
The six scalar functions $f_{1-6}$ depend solely on $|\vec{q}|$ and $\vec{q} \cdot \vec{n}$.
They are simply related to the commonly used longitudinal momentum fraction $x = k_1^+ / p^+$ and relative transverse momentum $\vec{k}_\perp = \vec{k}_{1\perp} - x \vec{p}_\perp$ as,
\begin{align}
    q^2 &= \frac{k_\perp^2 + m^2}{4x(1-x)}-m^2,\quad \vec{q} \cdot \vec{n} = \sqrt{\frac{k_\perp^2 + m^2}{x(1-x)}}\left(\frac{1}{2}-x\right)\,.
\end{align}
Meson exchange is one of the standard approaches for quantitatively describing the nucleon force in low-energy nuclear physics. Here, we employ a one-boson-exchange (OBE) kernel first proposed in Ref.~\cite{Machleidt:1987hj} to describe the NN interaction within the deuteron. The kernel includes pseudoscalar ($\pi, \eta$), scalar ($\sigma, \delta$), and vector ($\rho, \omega$) channels. It is a reparametrization of the full Bonn interaction, a high-precision potential that includes both OBE and multi-meson contributions. For the crucial mesons ($\pi, \rho, \omega$), the parameters are taken directly from the full Bonn interaction, whereas those for the remaining mesons ($\eta, \sigma, \delta$) are adjusted to reproduce the deuteron properties and low-energy NN scattering parameters. The $\eta$ and $\sigma$ mesons are newly introduced to compensate for the omission of multi-meson contributions. This procedure preserves the accurate description of the vast majority of the low-energy scattering phase shift data and the deuteron properties while simplifying the calculation. For each nucleon-meson coupling vertex, a form factor is included to account for the finite size of hadrons. The complete parameter set can be found in Ref.~\cite{Carbonell:1995yi}.

Figure~\ref{fig:Deuteron-LFWF} shows the functions $f_{1-6}$ obtained by numerically solving the light-front Schr\"{o}dinger equation with the OBE interaction kernel within the weak-binding approximation~\cite{Carbonell:1995yi}. At low momentum $q \lesssim 0.5~\text{GeV}$, the wave function is dominated by $f_1$ and $f_2$, while $f_5$ becomes comparable in magnitude for $q \gtrsim 0.5~\text{GeV}$. In the non-relativistic limit $q \ll m$, the functions $f_{3-6}$ are negligible, and $f_{1}$, $f_2$ reduce to the conventional non-relativistic S- and D-wave components. The LFWFs data are available in Ref.~\cite{Carbonell:1995yi} and satisfy the normalization condition,
\begin{gather}
    \frac{1}{3} \sum_{\lambda,\sigma_1,\sigma_2} \int_0^1 \frac{\text{d}x}{2x(1-x)} \int \frac{\text{d}^2k_\perp}{(2\pi)^3} |\Psi^\lambda_{\sigma_2\sigma_1}(x,\vec{k}_{\perp})|^2 = 1.
\end{gather}

\begin{figure*}
    \centering
    \includegraphics[scale=0.465]{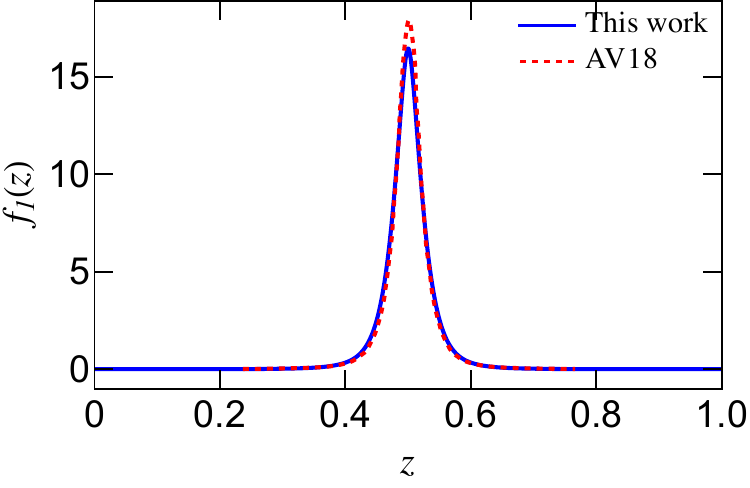}
    \includegraphics[scale=0.465]{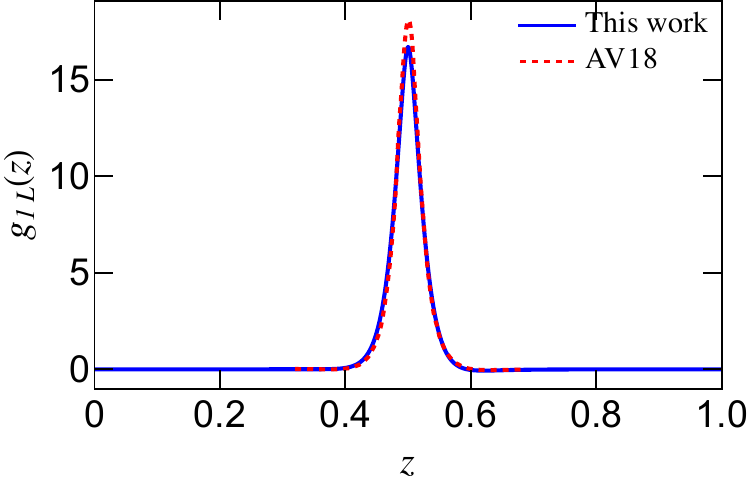}
    \includegraphics[scale=0.465]{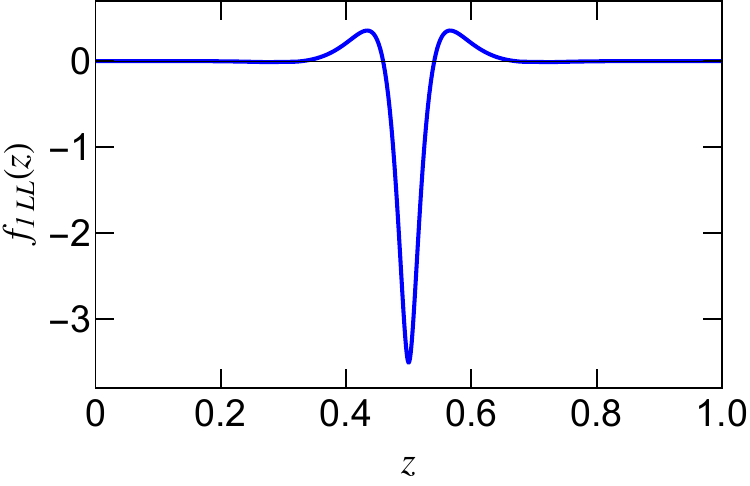}
    \caption{The unpolarized $f_{1}$ (left), longitudinally polarized $g_{1L}$ (middle) and tensor polarized $f_{1LL}$ (right) LMDs of deuteron. The unpolarized and longitudinally polarized LMDs are compared with AV18 potential model deuteron wavefunctions~\cite{Fornetti:2023gvf,Cosyn:2020kwu}.}
    \label{fig:LMDs}
\end{figure*}

\section{Longitudinal momentum distributions}\label{sec:observables}
The deuteron LMDs which describe the probability of finding a constituent carrying a fractional longitudinal momentum of deuteron $z$ can be obtained via the correlator~\cite{Mulders:1995dh},
\begin{eqnarray}\label{eq:LMD_correlator}
    \Phi_{\Lambda}^{{ij}}(z)=\int\frac{{\rm d}y^{-}}{2\pi}e^{ik^{+}.y^{-}}\langle P\Lambda|\bar{\psi}^{j}(0)\mathcal{U}_{[0,y^{-}]}\psi^{i}(y^{-})|P\Lambda\rangle\,.
\end{eqnarray}
where $P,~\Lambda$ denote the momentum and helicity of the deuteron, respectively. The fields, $\bar{\psi}$ and $\psi$ represent the fermionic field operators with Dirac indices $i,j$ and $\mathcal{U}_{[0,y^{-}]}$ is the gauge link which ensures the gauge invariance of the correlator. Within the light-cone gauge, $A^{+}=0$, the gauge link reduces to unity, i.e., $\mathcal{U}_{[0,y^{-}]}=1$. The partonic distributions can be extracted from the correlator by inserting the Gamma matrix as, $\Phi_{\Lambda}^{[\Gamma]}(z)=\frac{1}{2}\text{Tr}(\Gamma\Phi_{\Lambda}(z))$. From this correlator, the deuteron LMDs can be parametrized for different Dirac structures as follows~\cite{Bacchetta:2000jk},
\begin{eqnarray}\label{eq:LMDs_parametrization}
    \Phi_{\Lambda}^{[\gamma^{+}]}(z)&=&f_{1}(z)+S_{LL}f_{1LL}(z)\,,\nonumber \\
    \Phi_{\Lambda}^{[\gamma^{+}\gamma_{5}]}(z)&=&S_{L}g_{1L}(z)\,.
\end{eqnarray}
where $f_{1}(z)$ denotes the unpolarized LMD of deuteron, describing the probability of finding an unpolarized constituents carrying a fractional longitudinal momentum $z$ inside an unpolarized deuteron. The function $g_{1L}(z)$ represents the longitudinally polarized deuteron LMD, which characterizes the probability of a longitudinally polarized constituent carrying a momentum fraction $z$ within a longitudinally polarized deuteron. Finally, $f_{1LL}(z)$ corresponds to the tensor-polarized LMDs of the deuteron, describing the distribution of unpolarized constituents inside a tensor-polarized deuteron. $S_{L}$ is the longitudinal polarization and $S_{LL}=(3\Lambda^2-2)(\frac{1}{6}-\frac{1}{2}S_{L}^{2})$.
Note that for the longitudinally polarized ($\Lambda=0$)
deuteron, we have $S_{L}=1$,
and for transversely polarized deuteron ($\Lambda=\pm1$), $S_{L}=0$~\cite{Ninomiya:2017ggn}.
The trace of the correlator can be expressed via the LFWFs,
\begin{align}
    &\Phi^{[\gamma^+]}_\lambda = \sum_{\sigma_1,\sigma_2} \int \frac{\text{d}^2 k_\perp}{(2\pi)^3} \frac{\Psi^{\lambda*}_{\sigma_2\sigma_1}(z, \vec{k}_\perp) \Psi^\lambda_{\sigma_2\sigma_1}(z, \vec{k}_\perp)}{2z(1-z)},\\
    &\Phi^{[\gamma^+ \gamma_5]}_\lambda = \notag\\
    &\sum_{\sigma_1,\sigma_1^\prime,\sigma_2} \int \frac{\text{d}^2 k_\perp}{(2\pi)^3} \frac{\Psi^{\lambda*}_{\sigma_2\sigma_1^\prime}(z, \vec{k}_\perp) \mathcal A_{\sigma_1^\prime\sigma_1} \Psi^\lambda_{\sigma_2\sigma_1}(z, \vec{k}_\perp)}{2z(1-z)}.
\end{align}
where $\mathcal A = R_M^\dagger \sigma_3 R_M $ is a rotation matrix connecting the light-front and canonical spinors, with $R_M$ the Melosh rotation~\cite{Melosh:1974cu}.
Equation~\eqref{eq:deuteron-LFWF-spinor-structure} is formulated in terms of canonical spin basis. In the zero transverse momentum limit, $\vec{k}_{1\perp} = 0$, one has $\mathcal A_{\sigma^\prime\sigma} = \operatorname{sign}(\sigma)\delta_{\sigma^\prime\sigma}$, recovering the familiar light-front expression.

The deuteron unpolarized, longitudinally polarized and tensor polarized LMDs further can be expressed in terms of the number densities of it's constituents as follows,
\begin{eqnarray}
    f_{1}(z)&=&\frac{1}{3}[\Phi_{0}^{[\gamma^{+}]}(z)+\Phi_{+1}^{[\gamma^{+}]}(z)+\Phi_{-1}^{[\gamma^{+}]}(z)]\,,\nonumber\\
    g_{1L}(z)&=& \Phi_{+1}^{[\gamma^{+}\gamma_{5}]}(z)\,,\nonumber\\
    f_{1LL}(z)&=& \frac{1}{2}\left[2\Phi_{0}^{[\gamma^{+}]}(z)-\Phi_{+1}^{[\gamma^{+}]}(z)-\Phi_{-1}^{[\gamma^{+}]}(z)\right]\,.
\end{eqnarray}
The deuteron structure functions encode its partonic substructure across different polarization states. The relation between the deuteron structure functions and parton distribution functions~\cite{Hoodbhoy:1988am} is given as follows,
\begin{eqnarray}\label{eq:deuteron structure functions}
    \text{SF}^{D}(x,Q^2)=x\sum_{i}e_{i}^{2}\mathcal{F}_{i}^{D}(x,Q^2)\,.
\end{eqnarray}
where $x$ denotes the longitudinal momentum fraction carried by a parton inside the nucleon, and $i$ labels the quark flavor with electric charge $e_{i}$ (e.g., $e_{u}=+2/3,~e_{d}=-1/3$). The function $\mathcal{F}_{i}^{D}$ represents the parton distribution of the deuteron, which can be further factorized as a convolution of the deuteron LMDs and the quark distributions inside the nucleon, as follows~\cite{Akulinichev:1985ij,Fornetti:2023gvf,Kaur:2025css},
\begin{eqnarray}\label{eq:deuteron pdfs}
    \mathcal{F}_{i}^{D}(x,Q^2)=\frac{1}{2}\sum_{q}\int\frac{{\rm d}z}{z}\text{LMD}(z)\mathcal{F}_{i}^{N}\left(\frac{x}{z},Q^2\right)\,.
\end{eqnarray}
where $\mathcal{F}_{i}^{N}$ is the quark parton distributions inside a nucleon.

\begin{figure*}[t]
    \centering
    \includegraphics[scale=0.32]{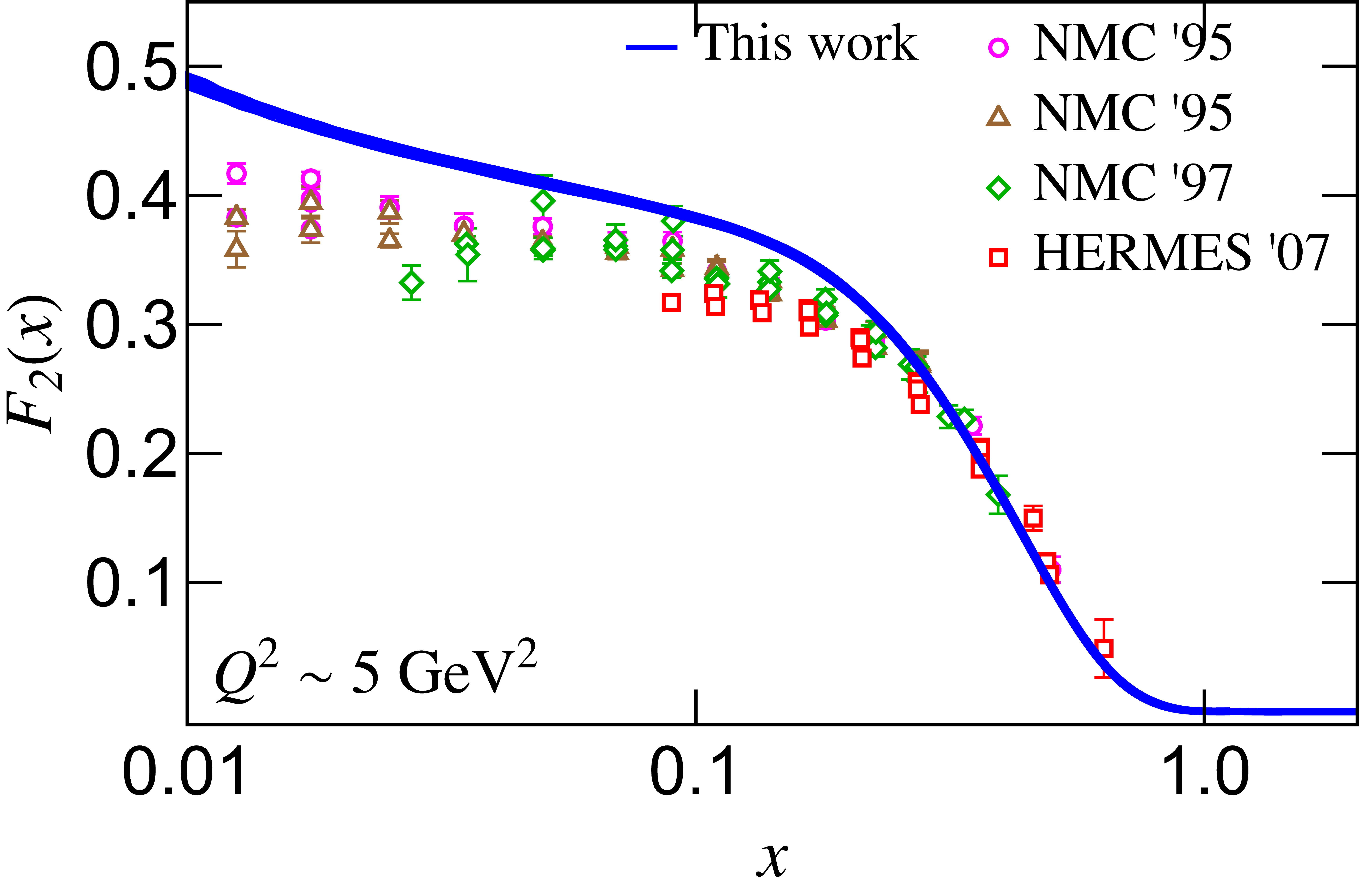}
    \includegraphics[scale=0.34]{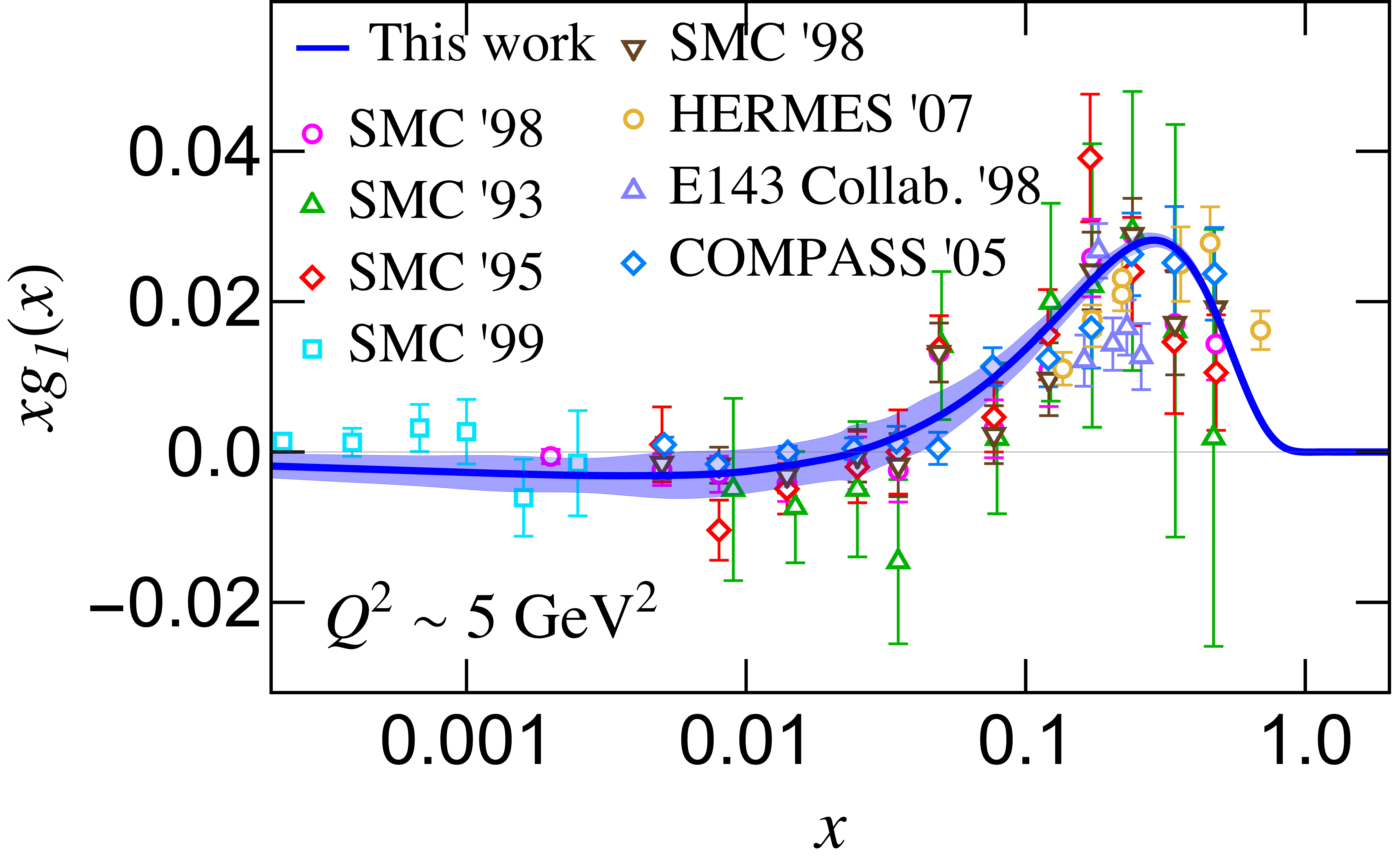}
    \includegraphics[scale=0.345]{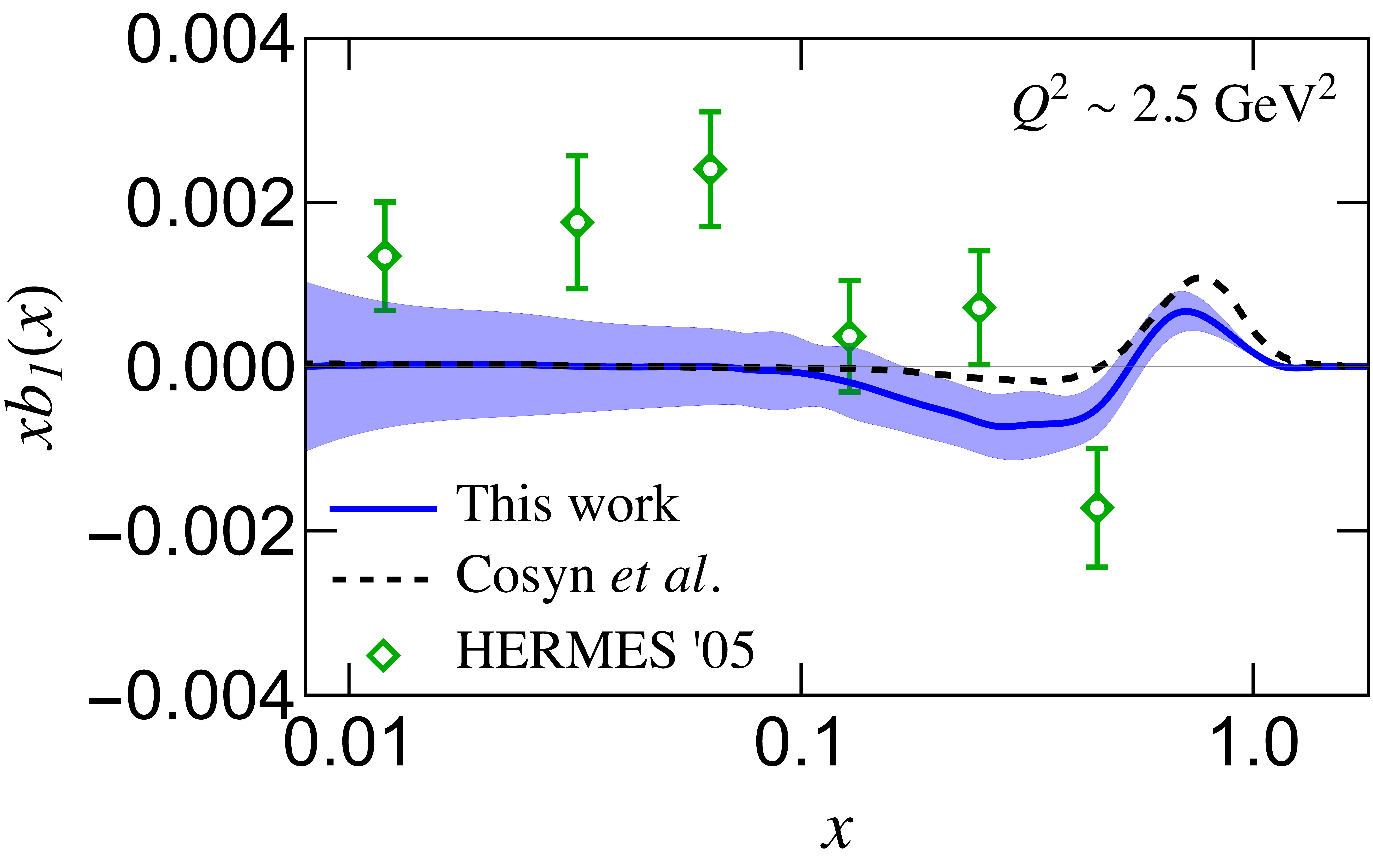}
    \caption{Comparison of unpolarized $F_{2}$ (left), helicity $g_{1}$ (middle) and tensor polarized $b_{1}$ (right) deuteron structure functions with available experimental data.}
    \label{fig:SFs}
\end{figure*}
\section{Results and Discussion}\label{sec:results and discussions}
In Fig.~\ref{fig:LMDs}, we present the deuteron LMDs and compare them with the momentum distributions obtained from the AV18 potential~\cite{Wiringa:1994wb,Fornetti:2023gvf,Cosyn:2020kwu}. Although minor differences exist in the peak height and width between our model (solid blue line) and the AV18 potential (red dashed line), both distributions integrate to unity reflecting the proper normalization of the deuteron wave functions. The left, middle, and right panels correspond to the unpolarized, longitudinally polarized, and tensor-polarized deuteron LMDs, respectively. Note that the unpolarized and tensor-polarized LMDs obey the momentum sum rules~\cite{Ninomiya:2017ggn,Kaur:2020emh}: The unpolarized distribution satisfies $\int{\rm d}zf_{1}(z)=1$, which reflects the normalization condition that the total longitudinal momentum fraction carried by the constituents of the deuteron sums to unity. Whereas, the tensor-polarized distribution obeys $\int{\rm d}zf_{1LL}(z)=0$, indicating that the tensor polarization does not contribute to the net momentum when integrated over all momentum fractions. As a consequence of this constraint, the function $f_{1LL}(z)$ must change sign within the physical region, implying the existence of at least one node (zero-crossing point). This behavior is clearly illustrated in the right panel, where the distribution exhibits both positive and negative regions to ensure the sum rule is satisfied.

Note that the deuteron LMDs are significantly narrower than the quark PDFs inside the nucleon because the deuteron is a weakly bound nuclear system with small nucleon Fermi motion. Consequently, the proton and neutron each carry nearly half of the deuteron longitudinal momentum, leading to distributions sharply peaked around the equal-momentum configuration, $z\sim0.5$. In contrast, the quark PDFs inside the nucleon are governed by relativistic quark-gluon dynamics and QCD interactions, which generate broad momentum distributions over a wide kinematic range. Therefore, the deuteron mainly introduces a mild smearing effect on the nucleon structure functions within the convolution formalism~\cite{Ehlers:2014jpa}.

In Fig.~\ref{fig:SFs}, we compare the unpolarized, longitudinally polarized, and tensor-polarized deuteron structure functions with the available experimental data. The left and middle panels display the unpolarized structure function, $F_{2}(x)$, and the longitudinally polarized structure function, $g_{1}(x)$, respectively. Both are evaluated within a convolution framework: $F_{2}(x)$ is obtained by combining the unpolarized deuteron LMD with the unpolarized quark PDFs, while $g_{1}(x)$ is calculated using the longitudinally polarized deuteron LMD together with the helicity-dependent quark PDFs of nucleons. This formalism effectively incorporates the internal momentum distribution of nucleons within the deuteron.
Our results for $F_{2}(x)$ are compared with measurements from the New Muon Collaboration (NMC)~\cite{NewMuon:1995zeb,NewMuon:1996fwh} and the HERMES~\cite{Gabbert:2008zz} data sets, while those for $g_{1}(x)$ are compared with data from the Spin Muon Collaboration (SMC)~\cite{SpinMuon:1998eqa,SpinMuon:1993gcv,SpinMuon:1999udj,SpinMuon:1995svc}, HERMES~\cite{HERMES:2006jyl}, and COMPASS~\cite{COMPASS:2005xxc} measurements. The deviation of the deuteron structure function $F_{2}(x)$ from the experimental data in the small-$x$ region can be attributed to limitations of the convolution framework, where nuclear effects are encoded only through LMDs folded with nucleon PDFs. In this region, the impulse approximation becomes insufficient, as additional nuclear phenomena such as shadowing, off-shell corrections, and medium modifications are absent in the input nucleon PDFs which can play a significant role~\cite{Kulagin:2004ie}. Consequently, this simplified factorization fails to fully capture the nuclear dynamics, leading to a mismatch with the data at low $x$. Overall, the calculated structure functions are qualitatively in agreement with the experimental data from moderate to large $x$-region, indicating that the adopted framework provides a reliable description of the deuteron structure in the considered kinematical region.

In the right panel of Fig.~\ref{fig:SFs}, we present a comparison of the tensor-polarized deuteron structure function. This is obtained by convoluting the tensor-polarized deuteron LMD with the unpolarized quark PDFs of the nucleons. The results are compared with the available data from HERMES~\cite{HERMES:2005pon} as well as with  another convolution model of W. Cosyn {\it et al}.~\cite{Cosyn:2017fbo}, evaluated at the average HERMES scale $Q^2 \sim 2.5~\text{GeV}^2$. Furthermore, the overall sign and magnitude of our predictions are consistent with those obtained in convolution formalism by W. Cosyn {\it et al}. However, the overall sign of our predictions is opposite to the rather old estimates reported in Ref.~\cite{Khan:1991qk} and to recent calculations based on light-front holography~\cite{Kaur:2025css} where the authors incorporated the hidden-color degrees of freedom, describing the deuteron as an effective superposition of singlet-singlet and octet-octet color clusters that extends beyond the conventional proton-neutron picture, although both works employ a convolution formalism. We find that our predictions for $b_{1}(x)$ are significantly smaller than the HERMES data at $x<0.1$, indicating that conventional deuteron models cannot account for the discrepancy and that additional hadronic dynamics beyond the standard convolution framework may be required~\cite{Cosyn:2017fbo,Kumano:2024fpr}. It is also interesting to see the nonzero contributions even at larger $x>1$. On the other hand, it has been suggested that $b_{1}$ may receive contributions from novel hadronic mechanisms, such as hidden color~\cite{Miller:2013hla,Kaur:2025css}. The upcoming Jefferson Lab experiment will measure $b_{1}(x)$ at medium $x$~\cite{ProposalJLab2023}, and our results can be regarded as theoretical predictions for this measurement.

\begin{figure}
    \centering
    \includegraphics[width=0.45\textwidth]{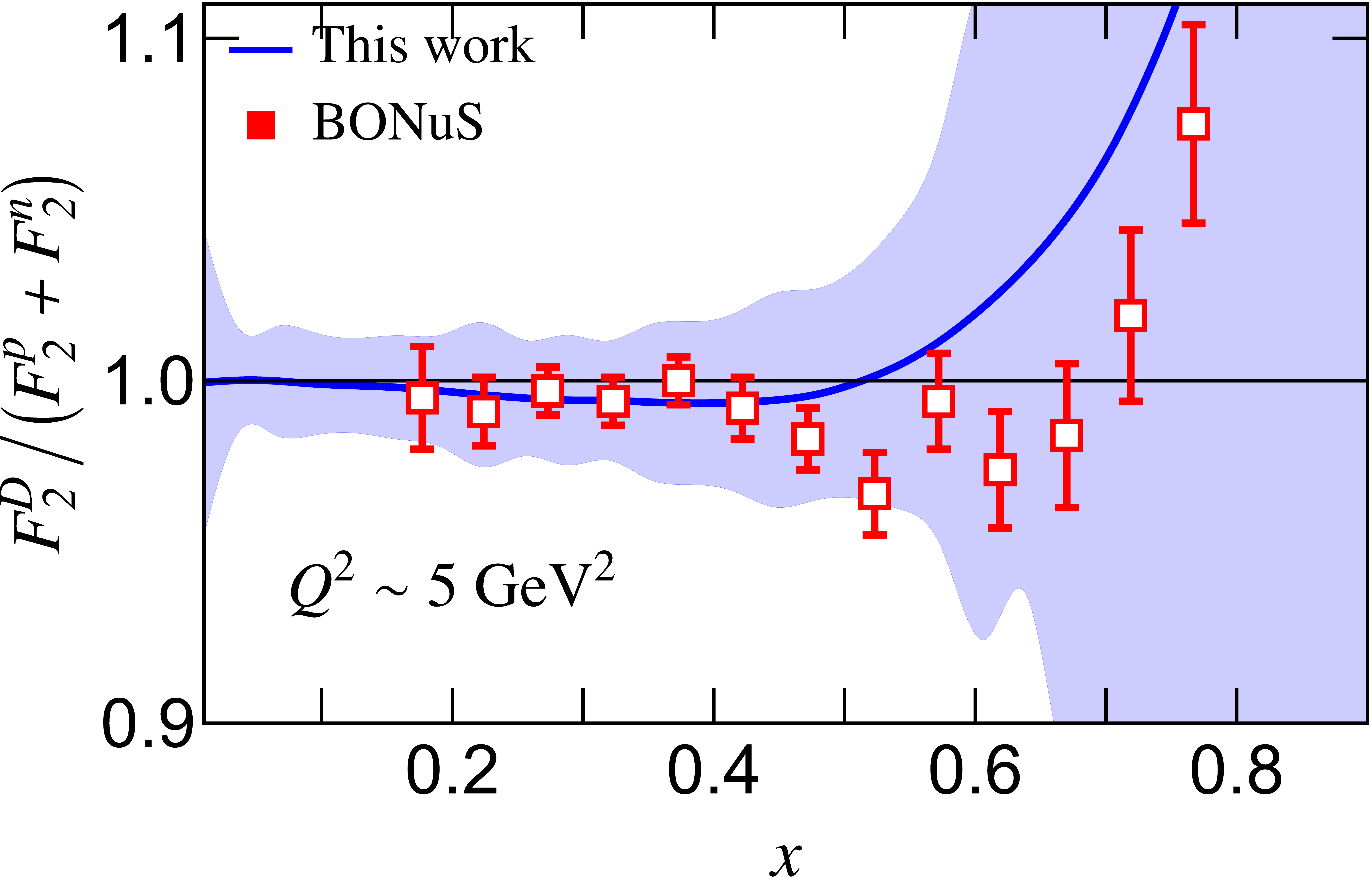}
    \caption{The ratio of the deuteron unpolarized structure function $F_2^D(x)$ to the sum of the free proton and neutron structure functions. Our calculation is compared with the BONuS data (red squares)~\cite{Griffioen:2015hxa}.}
    \label{fig:EMC}
\end{figure}

Finally, we present the $x$ dependence of EMC effect of the deuteron in Fig.~\ref{fig:EMC}. Our results are in qualitative agreement with experimental data: the ratio approaches 1 in the small-$x$ region, decreases in the intermediate-$x$ region, and exhibits an enhancement in the large-$x$ region. The decrease in the intermediate region, identified as the EMC effect, is primarily attributed to nuclear binding, while the enhancement for $x > 0.8$ arises from Fermi motion~\cite{Kulagin:2004ie}. However, compared with the BONuS data~\cite{Griffioen:2015hxa}, our calculation enters the suppression region at a lower $x$ value, and a similar offset is observed in the enhancement region. When uncertainties are taken into account, the theoretical results are consistent with the experimental data. Nevertheless, the theoretical errors remain substantial, primarily due to the uncertainties propagated from the global fit results used in the convolution. Systematic uncertainties intrinsic to the theoretical model itself are also expected, although they are difficult to estimate. To better understand the discrepancy between experiment and theory, more systematic theoretical approaches, such as chiral effective theory, as well as more precise PDF data are required.

\section{Conclusion}
In this work, we utilize relativistic deuteron light-front wave functions obtained from
a realistic nucleon-nucleon interaction to compute the LMDs and corresponding structure functions within the standard convolution framework. Using the NNPDF parton distributions as input, we find that the calculated unpolarized and longitudinally polarized structure functions provide a good description of the experimental data. However, the predicted tensor-polarized structure function $b_1$ remains significantly smaller than the HERMES measurements, suggesting that the standard convolution formalism may not be sufficient to fully interpret the data.
The existing HERMES experimental data for the tensor-polarized structure function are subject to large uncertainties, motivating more precise measurements in future experiments. In particular, improved studies of the tensor-polarized spin-1 deuteron are planned at Jefferson Lab, while complementary information may be obtained from proton–deuteron Drell–Yan processes at Fermilab. Polarized deuteron beams at NICA and future EIC facilities are also expected to provide broader opportunities to investigate the spin structure of the deuteron.

We further investigate the deuteron EMC effect by comparing the unpolarized deuteron structure function with the corresponding free proton and neutron contributions to examine the influence of the deuteron wave function on parton distributions. We employ the NNPDF data set to calculate the free proton structure function, while the neutron contribution is constructed assuming charge and isospin symmetry. Our analysis reproduces the qualitative behavior observed in the BONuS data, although noticeable differences remain in the position of the EMC suppression region.

Within the current two-nucleon picture, the theoretical calculations are unable to simultaneously describe all experimental data. The remaining discrepancies may point to dynamics beyond the conventional hadronic description and could provide sensitivity to the underlying QCD structure of the deuteron~\cite{Bakker:2014cua}. Nevertheless, several effects are still absent from the present framework. There remains a possibility of reducing the disagreement by incorporating higher-twist effects as well as non-nucleonic degrees of freedom into the present model~\cite{Ericson:1983um}. A natural extension of this work is to incorporate higher Fock-state components, such as pionic contributions, that go beyond the two-nucleon picture of the deuteron.

\section{Acknowledgment}
The authors wish to thank C. Mondal, S. Kaur, S. Xu, X. Zhao, C. Weiss, C.-R. Ji, S. Yang, L. Han, and C.P. Yuan for illuminating discussions carried out during the course of this work. X.C. is supported by the National Natural Science Foundation of
China (NSFC) under Grant No.~125B2111. Y.L. is supported by the Chinese Academy of Sciences under Grant No.~YSBR-101.

\end{document}